\documentclass[pra,twocolumn, showpacs]{revtex4}

\usepackage{graphicx}

\begin{document}

\def\kket{\rangle \mskip -3mu \rangle}
\def\bbra{\langle \mskip -3mu \langle}

\def\ket{\rangle}
\def\bra{\langle}

\def\pard{\partial}

\def\sinh{{\rm sinh}}
\def\sgn{{\rm sgn}}

\def\t{t_\mathrm{h}}
\def\tha{\hat t}
\def\alp{\alpha}
\def\del{\delta}
\def\Del{\Delta}
\def\eps{\epsilon}
\def\gam{\gamma}
\def\sig{\sigma}
\def\kap{\kappa}
\def\lam{\lambda}
\def\ome{\omega}
\def\Ome{\Omega}

\def\th{\theta}
\def\vphi{\varphi}

\def\Gam{\Gamma}
\def\Ome{\Omega}

\def\kav{{\bar k}}

\def\abf{{\bf a}}
\def\cbf{{\bf c}}
\def\dbf{{\bf d}}
\def\gbf{{\bf g}}
\def\kbf{{\bf k}}
\def\lbf{{\bf l}}
\def\nbf{{\bf n}}
\def\pbf{{\bf p}}
\def\qbf{{\bf q}}
\def\rbf{{\bf r}}
\def\ubf{{\bf u}}
\def\vbf{{\bf v}}
\def\xbf{{\bf x}}
\def\Cbf{{\bf C}}
\def\Dbf{{\bf D}}
\def\Kbf{{\bf K}}
\def\Pbf{{\bf P}}
\def\Qbf{{\bf Q}}

\def\omet{{\tilde \ome}}
\def\gammat{{\tilde \gamma}}
\def\Ft{{\tilde F}}
\def\It{{\tilde I}}
\def\ut{{\tilde u}}
\def\bt{{\tilde b}}
\def\Vt{{\tilde V}}
\def\xt{{\tilde x}}

\def\ph{{\hat p}}

\def\wt{{\tilde w}}
\def\xit{{\tilde \xi}}
\def\phit{{\tilde \phi}}
\def\rhot{{\tilde \rho}}

\def\Cb{{\bar C}}
\def\Nb{{\bar N}}
\def\Ab{{\bar A}}
\def\Db{{\bar D}}
\def\gb{{\bar g}}
\def\nb{{\bar n}}
\def\bb{{\bar b}}
\def\Pib{{\bar \Pi}}
\def\rhob{{\bar \rho}}
\def\phib{{\bar \phi}}
\def\psib{{\bar \psi}}
\def\omeb{{\bar \ome}}

\def\Sh{{\hat S}}
\def\Wh{{\hat W}}

\def\SS{I}
\def\psiw{{\xi}}
\def\tI{{g}}

\def\Ep#1{Eq.~(\ref{#1})}
\def\Eqs#1{Eqs.~(\ref{#1})}
\def\EQN#1{\label{#1}}

\newcommand{\beqa}{\begin{eqnarray}}
\newcommand{\eeqa}{\end{eqnarray}}


\title{Existence and nonexistence of an intrinsic tunneling time}
\author{Gonzalo Ordonez}
\affiliation{Department of Physics, Butler University, 4600 Sunset Ave., 
Indianapolis, IN 46208, USA.}

\author{ Naomichi Hatano}
\affiliation{Institute of Industrial Science, University of Tokyo,
Komaba 4-6-1, Meguro, Tokyo 153-8505, Japan}

\date{\today}


\begin{abstract}
Using a time operator, we define a tunneling time for a particle going through a barrier. This tunneling time is the average of the phase time introduced by other authors. In addition to 
the delay time caused by the  resonances over the barrier, the present tunneling time is also affected by the branch  point  at the edge of the energy continuum. We find that when the particle energy is 
near the branch point, the tunneling time becomes strongly dependent on the 
width of  the incoming wave packet, which implies that there is no intrinsic tunneling 
time. This effect is related to the quantum uncertainty in the particle's momentum.
\end{abstract}


\maketitle

\section{Introduction}
The definition of tunneling time --- the time it takes a particle to tunnel 
through a 
potential barrier --- or even whether it can be defined or not, has been a much 
debated problem and is still a controversial one of fundamental quantum 
mechanics~\cite{Bohm,Wigner,Buttiker82,Buttiker85,Leavens,Landauer92,Imafuku95,Imafuku97,Hara00,Hara03,Hauge,Olkhovsky,Olkhovsky2, Landauer94,Razavy,Yamada0,Buttiker, Landauer, Davis, Yamada}. 
In this paper we address the question: Is there an intrinsic tunneling time? We  present a definition of tunneling  time on the basis of a time 
operator  canonically conjugate to the Hamiltonian~\cite{Olkhovsky, Kobe}. Our  
tunneling time consists of two contributions. The answer to the question 
is 
``Yes" for one contribution but ``No" for the other contribution.

The first contribution comes from the overlaps between the incoming wave and 
resonant states. It is basically a weighted sum over all resonance poles of the 
resonant lifetimes. We may say that the incoming wave splits into resonant 
channels of the tunneling barrier and spends the lifetime of each resonance 
before it tunnels out. For a particle represented by a spatially large wave 
packet, it is closely related to the phase time defined by Wigner, Smith, 
Pollak and Miller, and others~\cite{Wigner, Razavy,Smith,Pollak}. This is a dominant 
contribution to the tunneling time when the particle energy is near the 
resonance 
poles. \footnote{When the energy of the particle is larger than the height of the potential barrier ($E>V_0$), the particle does not tunnel anymore; it is transmitted. We may then formally  call the time it takes the particle to go through the barrier the transmission time for $E>V_0$. But since there is no algebraic difference between the cases $E<V_0$ and $E>V_0$, we hereafter tentatively still call this time the tunneling time even for $E>V_0$. Note that for energies close to $V_0$ (but below $V_0$, \textit{i.e.}, in the true tunneling range) the resonance poles do have an influence on the tunneling process, because resonances have a finite width.} It gives a tunneling time (as a function of the particle energy) that is 
independent of the width of the incoming wave packet. In other words, it gives 
an intrinsic tunneling time, which depends only on the resonance poles of the 
barrier.

The second contribution appears when the particle energy is near a branch point. 
In contrast to the first contribution, it is strongly dependent on the width of 
the incoming wave packet. This makes a universal definition of the tunneling time \textit{impossible} near 
a branch point. 

In short, our main point is that when the energy of the incoming particle is 
near a resonance pole of the tunneling barrier, an intrinsic tunneling time 
does exist, but when the energy is near the branch point there is no intrinsic 
tunneling time.  This is understandable; while the resonance poles yield the 
Markovian dynamics (exponential decay), the branch point yields non-Markovian 
dynamics (\textit{i.e.}, power-law decay) with no characteristic time or length scales, 
which cause deviations from exponential decay for both  long time 
scales~\cite{Khalfin} and short time scales~\cite{Misra,Petrosky}.
In the following, we present general arguments to support our claim and present 
numerical results for a square-barrier model. 

\section{Time operator and age}
Our argument starts with the time operator~\cite{Olkhovsky, Kobe}
\begin{eqnarray} \label{timeop}
\tha =  i \frac{\pard}{\pard H'}
 \end{eqnarray}
in units with $\hbar=1$. Here $H'$ is the part of the Hamiltonian associated 
with a 
continuous spectrum, or the Hamiltonian excluding the bound 
states of the particle,
\begin{eqnarray} \label{cset}
 H' =   \sum_{\alpha} \int_{-\infty}^\infty \frac{dk}{2\pi} \, |E_k^{\alpha} 
\ket E_k^{\alpha} \bra E_k^{\alpha}|  
  \end{eqnarray}
with $E_k^{\alpha}$ denoting the dispersion relation of a mode $\alpha$ with 
wave number $k$ of free propagation. 
In terms of the eigenstates of the Hamiltonian we have
\begin{eqnarray}
 \tha = \sum_{\alpha}  \int_{-\infty}^\infty \frac{dk}{2\pi} \, |E_k^{\alpha} 
\ket i \frac{\pard}{\pard E_k} \bra E_k^{\alpha}|.
\end{eqnarray}

The time operator~(\ref{timeop}) satisfies the commutation relations
$[\tha, H'] = i$
and $[\tha, H_b] = 0$, where $H_b$ is the part of $H$ that includes the bound 
states. 
These commutation relations  give
$e^{iHt}\tha e^{-iHt}  = \tha + t$,
so that the time evolution just adds time $t$ to the time operator. This 
property allows us to interpret the time operator as giving the ``age'' of a 
state~\cite{Misra2}. We define the average age of a normalized state $|\psi\ket$  
as 
\begin{eqnarray}
\bra \tha \ket_\psi = \bra \psi|\tha|\psi\ket.
 \end{eqnarray}

Since the energy is bounded from below, 
the time operator is not self-adjoint~\cite{Pauli,OR07}, \textit{i.e.},
$\bra A|\tha|B\ket \ne \bra B|\tha|A\ket^*$.
This means that the age of a given state at $t=0$ can be complex. 
However, we will consider the age {\it difference} between 
incoming and outgoing states of the particle. In momentum representation the states we consider will differ only by a phase factor. As a result, the age difference will be real, despite the ages being complex. 

The difference in age between two states is given by
\begin{eqnarray}\label{agediff}
 t_{\psi_2,\psi_1} = \bra \tha \ket_{\psi_2} -  \bra \tha \ket_{\psi_1}.
 \end{eqnarray}
Keeping in mind an experimental scenario where  the particle has an average 
positive velocity (moving from left to right) and tunnels through a  potential 
barrier, we define  the initial state $\psi_1$ as a state where the particle is 
known to be on the left side of the barrier, and the final state $\psi_2$ as a 
state  where  the particle is known to be on the right side of the barrier, with 
both $\psi_1$ and $\psi_2$ giving the same average velocity.  Our postulate is 
that $ t_{\psi_2,\psi_1}$ will then give an average of the time it takes the 
particle to move from the left of the barrier to the right.
Note that $\psi_2$ is {\it not} the time-evolved state $\psi_1(t)$. If $\psi_2$ 
were taken as $\psi_2 = \psi_1(t)$, due to 
the relation $e^{iHt}\tha e^{-iHt}  = \tha + t$,
the age difference $t_{\psi_1(t),\psi_1(0)}$ would simply give $t$.

\section{General form of the age difference}
Let us consider a general one-dimensional system with a symmetric potential 
barrier centered at 
$x=0$.
In position representation, the stationary eigenstates $|E_k\ket$ (giving the 
eigenvalue continuum) of the Hamiltonian have the form
\begin{equation}\label{Ek}
  \bra x|E_k\ket
  =
  \left\{ \begin{array}{ll}
        T(k) e^{ik(x-a)} , & x\ge a/2,\\
        B_k(x),  &  -a/2 \le x \le a/2,\\
       e^{ikx} + R(k) e^{-ik(x+a)}  , & x\le -a/2,
        \end{array} \right.
 \end{equation}
where $a$ is the width of the barrier, $R$ is the reflection coefficient, $T$ is 
the transmission coefficient, and $B_k(x)$ is the wave function inside the 
barrier. 

We  will use   the symmetric ($\alpha=+$) and anti-symmetric ($\alpha=-$) modes 
of the stationary states:
\begin{equation}\label{Ekp}
  \bra x|E_k^\pm \ket
  = \frac{1}{2} 
  \left\{ \begin{array}{ll}
        \pm e^{-ikx} \pm F_\pm(k) e^{ikx}, & x\ge a/2,\\
        B_k(x) \pm B_k(-x) , &  -a/2 \le x \le a/2,\\
       e^{ikx} + F_\pm(k) e^{-ikx} , & x\le -a/2.
        \end{array} \right.
 \end{equation}
The factor $1/2$ accounts for the normalization of the states $|E_k^\pm \ket$ 
and the double-counting of positive and negative $k$  in Eq.~(\ref{cset}).  The 
coefficient
\begin{equation}
 F_\pm(k) \equiv (R(k)\pm T(k))e^{-ika}
  \end{equation}
is the scattering amplitude for the symmetric or antisymmetric outgoing waves, respectively.
Since the incoming flux of $e^{-ik|x|}$ and the outgoing flux of $e^{ik|x|}$ should be equal for the stationary states, the scattering 
amplitude must have modulo $1$, \textit{i.e.}, $|F_\pm(k)|^2 = 1$ for real $k$.

In position representation, we set the initial and final states as
\begin{eqnarray}\label{xpsi1}
 \bra x|\psi_1\ket&=&\exp(ik_0 x)/ \sqrt{L_0} \quad {\rm for\,} -L_0-a/2 \le x 
\le -
a/2,\nonumber\\
 \bra x|\psi_2\ket&=&\exp(ik_0 x)/ \sqrt{L_0}  \quad {\rm for\,} a/2 \le x \le 
L_0+a/2,
\end{eqnarray}
and $0$ for other $x$. These are truncated plane waves of width $L_0$. We have chosen truncated plane waves because we want to study the limiting situation when these states approach plane waves. This occurs when $L_0\gg k_0^{-1}$. Then both $\psi_1$ and $\psi_2$  approach plane waves with well-defined momentum $k_0$. We could use different functions such as Gaussians, but the functions above seem to be the simplest ones to consider.  Hereafter we will refer to the truncated plane waves (\ref{xpsi1}) as the wave packets. 
In momentum representation, we have 
\begin{eqnarray} \label{fg}
\bra k|\psi_1\ket &=& f^*(k-k_0), \\
\bra k|\psi_2\ket &=&  e^{-i(k-k_0)(L_0+a)} f^*(k-k_0),\nonumber
 \end{eqnarray}
where $f^*(k-k_0)$ is the Fourier transform of $\bra x|\psi_1\ket$, 
\begin{eqnarray} \label{fksq}
f^*(k-k_0) = \frac{1}{\sqrt{L_0}}e^{i(k-k_0)a/2} \frac{1-e^{i(k-k_0)L_0}}{-i(k-
k_0)}.
 \end{eqnarray}

 In the end, we will take the limit $L_0\to\infty$. This limit will turn out to be unique for some terms of the age difference, but non-unique for other terms, because it will depend on other parameters such as  the momentum of the incoming particle.  We will judge the ``intrinsicness'' of the tunneling time by seeing whether it has a well-defined limit or not when the size of the wave packets tend to infinity, approaching plane waves with fixed momentum. 
 
 We  will calculate the age difference~(\ref{agediff}) as
$t_{\psi_2,\psi_1} = \bra\psi_2|\tha|\psi_2\ket - 
\bra\psi_1|\tha|\psi_1\ket$, where
\begin{eqnarray} \label{psitpsi}
\bra\psi_j|\tha|\psi_j\ket =  \sum_{\alpha=\pm}  \int_{-
\infty}^\infty \frac{dk}{2\pi} \bra\psi_j |E_k^{\alpha} 
\ket i \frac{\pard}{\pard E_k} \bra E_k^{\alpha}|\psi_j\ket.
\end{eqnarray}
 The age difference may be separated into the age 
difference with no barrier, and a delay $\Delta\tau$ due to the barrier, 
\begin{eqnarray}\label{aged}
t_{\psi_2,\psi_1} =   t_{\psi_2,\psi_1}^{(0)} +  \Delta\tau.
\end{eqnarray}
The delay time $\Delta\tau$ can be directly measurable, because  it is just the difference between the average arrival time of the particle with the barrier present and the average arrival time with no barrier present.

After some algebra (see Appendix \ref{AppB}), we obtain
\begin{eqnarray}\label{tprop}
  t_{\psi_2,\psi_1}^{(0)} =     (L_0+a)v^{-1},
 \end{eqnarray}
 where $v^{-1}$ is the average inverse group velocity $\pard k/\pard E_k$, 
\begin{eqnarray}\label{vav}
v(k_0)^{-1}  =      \int_{-\infty}^\infty \frac{dk}{2\pi}  |f(k-
k_0)|^2  \frac{\pard k}{\pard E_k},
\end{eqnarray}
and $L_0+a$ is the distance  from the middle region of the initial state
$\psi_1$ to that of the final state $\psi_2$. 
Some more algebra (see Appendix \ref{AppB}) allows us to split the delay time into two parts:
\begin{eqnarray}
\Delta\tau = \Delta\tau_A + \Delta\tau_B
\end{eqnarray}
with 
\begin{eqnarray}\label{ttun'}
 \Delta\tau_A (k_0) =     \frac{-i}{2} \sum_{\alpha=\pm}  \int_{-
\infty}^\infty \frac{dk}{2\pi}   |f(k-k_0)|^2  \frac{1}{F_\alpha(k)}
\frac{\pard F_\alpha(k)}{\pard E_k}  \nonumber\\
 \end{eqnarray}
 \begin{eqnarray}\label{ttun'b}
&&  \Delta\tau_B (k_0) =       \frac{i}{2} \int_{-\infty}^\infty 
\frac{dk}{2\pi} \left(f(-k+k_0) \frac{\pard}{\pard E_k} f(-k-k_0) \right. 
\nonumber\\
  &-&   \left. f(-k-k_0) \frac{\pard}{\pard E_k} f(-k+k_0)\right)  
\sum_{\alpha=\pm}  F_\alpha(k),
 \end{eqnarray}
 which are real, because $|F_\pm(k)|^2=1$,  $F_\pm(k)=F^*_\pm(-k)$, and $f(\kappa)=f^*(-\kappa)$ for real $k$ and $\kappa$. (In the calculation we neglected terms of order $1/L_0$). We will show in the next section that $\Delta\tau_A$ is related to the motion of the particle inside the barrier, \textit{i.e.}, the tunneling process.  On the other hand, $\Delta\tau_B$ is related to the motion of the particle outside the barrier. We can see this because it only involves the reflection coefficient $R$, which is proportional to $\sum_{\alpha=\pm} F_\alpha(k)$ in Eq.~(\ref{ttun'b}); see also the discussion below Eq.~(\ref{DADB'}).

\section{Tunneling time}
\begin{figure}[t]
\includegraphics[width=\columnwidth]{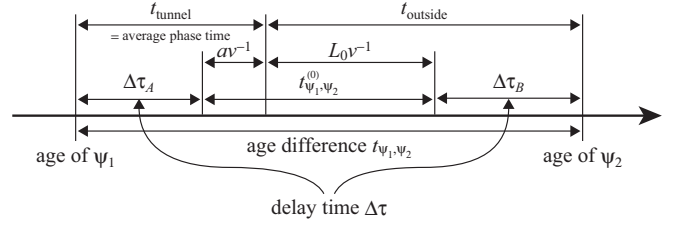}
\caption{Relation between the age difference, $t_{\rm tunnel}$, $t_{\rm outside}$, and the delay times $\Delta\tau_A$ and $\Delta\tau_B$} 
\label{ttimes}
\end{figure}

In this section we will define the tunneling time. We will first relate $\Delta\tau_A$ in Eq.~(\ref{ttun'}) to the 
phase time introduced by Wigner~\cite{Wigner}, Smith~\cite{Smith}, Pollak and Miller 
\cite{Pollak}, and others~\cite{Razavy}. This phase time is defined as
\begin{eqnarray}
 \tau_{\rm ph}(k) =     \Re\left( -i \frac{\pard}{\pard E_k} \ln T(k) \right) 
  \end{eqnarray}
 for a particle with the wave number $k$. Writing  $T(k) = |T(k)|e^{i\theta(k)}$, we 
have 
  $\tau_{\rm ph} =  \pard \theta\bigm/\pard E_k$.
To connect this to $\Delta\tau_A$, we write the amplitude 
as 
 $F_\pm(k) = \exp(i\th_\pm(k))$,
 where $\th_\pm(k)$ is a real phase.   Then 
\begin{eqnarray} \label{avphase}
 \Delta\tau_A =      \frac{1}{2} \sum_{\alpha=\pm}  \int_{-\infty}^\infty  
\frac{dk}{2\pi} |f(k-k_0)|^2    \frac{\pard}{\pard E_k}  \th_\alpha(k).
 \end{eqnarray}
 Moreover, since 
\begin{eqnarray}
 T &=& \frac{1}{2}e^{ika} \left(F_+-F_-\right)  = \frac{1}{2} e^{ika}\left(e^{i\th_+} - e^{i\th_-} \right) \\
   &=& i e^{i(\th_++\th_- + 2ka)/2} \sin{ \frac{\th_+ - \th_-}{2}},
  \end{eqnarray}
 we have  $\theta(k) = \pi/2 + (\th_+(k) +\th_-(k) + 2ka)/2$ and 
\begin{eqnarray}
\frac{1}{2} \sum_{\alpha=\pm} \th_\alpha(k) = \theta(k) - \pi/2 - ka,
 \end{eqnarray}
Therefore, from Eq.~(\ref{avphase}) we obtain
\begin{eqnarray} 
 \Delta\tau_A =      \int_{-\infty}^\infty  \frac{dk}{2\pi} |f(k-k_0)|^2    \frac{\pard}{\pard E_k}  (\theta(k) - ka).
 \end{eqnarray}
 Using Eq.~(\ref{vav}), we finally obtain 
\begin{eqnarray} \label{ttunredef}
\Delta \tau_A  =  \int_{-\infty}^\infty \frac{dk}{2\pi} |f(k-
k_0)|^2     \tau_{\rm ph}(k) -  a v^{-1}.
 \end{eqnarray}
The fist term on the right-hand side represents a weighted average of the phase time, where the weight is the square modulus of the Fourier component of the incident wave packet.
The second term $av^{-1}$ is the time it would take the particle to move through a distance $a$ (equal to the barrier width) by free propagation. 
The delay caused by the barrier is the difference between the tunneling time and the free-propagation time through the barrier region: $\Delta \tau_A  = t_{\rm tunnel} - a v^{-1}$.  Hence we identify the tunneling time with the average phase time as
\begin{eqnarray} \label{defttun}
t_{\rm tunnel} &=& av^{-1} + \Delta\tau_A\nonumber\\
&=& \int_{-\infty}^\infty \frac{dk}{2\pi} |f(k-k_0)|^2     \tau_{\rm ph}(k);
\end{eqnarray}
see Fig.~\ref{ttimes}.

\section{Evaluating the integrals}

In this section we will evaluate the integrals involved in the age difference, the delay time, and the tunneling time. 
We first note that in the momentum representation the time operator is given by 
\begin{eqnarray} \label{ddEk}
{\hat t} = i \frac{\pard}{\pard E_k}  = i \frac{m}{k} \frac{\pard}{\pard k},
 \end{eqnarray}
which diverges at $k=0$.  However, as we discuss now, this divergence is suppressed by  the term $\bra\psi_j |E_k^{\alpha} \ket$  in Eq.~(\ref{psitpsi}), which vanishes at $k=0$. In the $k$ representation the eigenstates (\ref{Ekp}) are given by 
\begin{eqnarray}
\bra\psi_1|E_k^\pm \ket &=& \frac{1}{2} \left(  \bra\psi_1|k\ket + F_\pm(k) \bra\psi_1|-k\ket \right), \nonumber\\
\bra\psi_2|E_k^\pm \ket &=& \pm\frac{1}{2} \left(  \bra\psi_2|-k\ket + F_\pm(k) \bra\psi_2|k\ket \right).
\end{eqnarray}
When $k\to 0$ we have $F_\pm(k) \to -1$, because when $k=0$ the particle cannot tunnel through the barrier as long as the barrier has a 
positive height and positive width. The particle is perfectly reflected. Hence $R(0)=-1$ and $T(0)=0$, which gives $F_\pm(0) = -1$.
(In Appendix \ref{AppSB} this is shown explicitly for a square-barrier potential). 

Thanks to this behavior of the scattering amplitude, we see that  $ \bra\psi_j|E_k^\pm\ket \to 0$ when $k\to 0$, for $j=1,2$. Moreover, the derivative  $\pard \bra E_k^\pm|  \psi_j\ket /\pard k$ is regular at $k=0$. Thus the vanishing $ \bra\psi_j|E_k^\pm\ket$ cancels the $1/k$ divergence at $k=0$ coming from Eq.~(\ref{ddEk}). This means that the point $k=0$ is not a true singularity in 
the age difference.

Since the integrand in the age difference is regular at $k=0$, we can replace $1/k$ by its principal part without changing the integration. This means that we can replace
\begin{eqnarray} \label{Attunsq3'}
 \frac{1}{ k} \to  \frac{1}{2}\left(\frac{1}{ k + i\eps} + \frac{1}{ k - i\eps} \right) 
   \end{eqnarray}
 with $\eps>0$ real (infinitesimal).  

 To evaluate the integrals in  Eqs.~(\ref{vav}) and~(\ref{defttun}),  we will also add an 
infinitesimal in the denominator of $|f(k-k_0)|^2$ as
$(k-k_0)^{-2}\to(k-k_0+i\eps)^{-2}$,
 which does not change the result  because $|f(k-k_0)|^2$ is regular at $k=k_0$. 
Similarly, we will add infinitesimals to the denominators of $f(\pm k\pm k_0)$ in the  integral of  $\Delta\tau_B$ in Eq.~(\ref{ttun'b}).

\begin{figure}[t]
\includegraphics[width=180pt, angle=90]{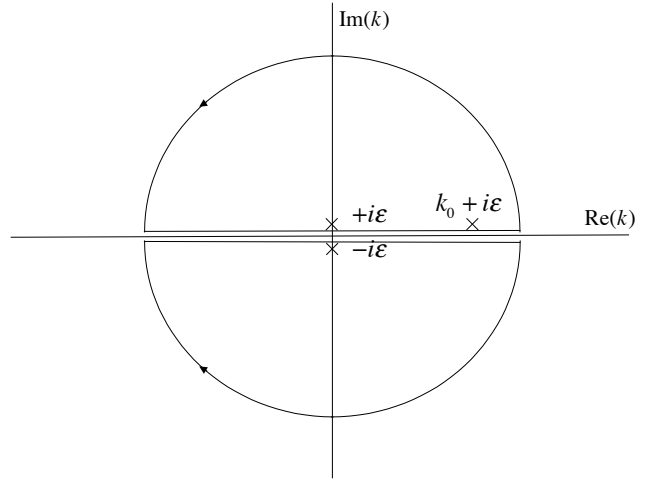}
\caption{Integration contours and poles of the age difference.} 
\label{Poles}
\end{figure}

 Let us consider first Eq.~(\ref{vav}), or  the average inverse velocity
\begin{eqnarray}
v^{-1}  &=&      \int_{-\infty}^\infty \frac{dk}{2\pi}  |f(k-k_0)|^2  \frac{\pard k}{\pard E_k} \nonumber\\
&=&  \int_{-\infty}^\infty \frac{dk}{2\pi}  |f(k-k_0)|^2  \frac{m}{k} \\
&=&  \frac{1}{L_0}  \int_{-\infty}^\infty \frac{dk}{2\pi} 
 \frac{\left|1-e^{i(k-k_0)L_0}\right|^2}{ (k-k_0)^2}\frac{m}{k}.\nonumber
\end{eqnarray}
 Expanding the absolute value squared and using Eq.~(\ref{Attunsq3'}), we obtain
\begin{eqnarray}
v^{-1}  &=&    \frac{1}{L_0}  \int_{-\infty}^\infty \frac{dk}{2\pi} 
 \frac{\left[1-e^{i(k-k_0)L_0}\right] + \left[1-e^{-i(k-k_0)L_0}\right]}{ (k-k_0 - i\eps)^2}\nonumber\\
 &\times&\frac{m}{2}\left(\frac{1}{k+i\eps}+ \frac{1}{k-i\eps}\right).
 \end{eqnarray}
   To evaluate the integral  we will close the contour with an infinite semicircle either on the upper or the lower half $k$-plane, depending on whether the integrand vanishes on the upper or lower infinite semicircle, respectively (see Fig.~\ref{Poles}).
For the first term in brackets in the numerator, we will close the contour on the upper half plane, and for the second term in brackets, we will close the contour in the lower half plane. Evaluating the residues at the double  pole $k=k_0 + i\eps$ and the poles $k=\pm i\eps$, we obtain
\begin{eqnarray} \label{Avinv}
v^{-1}  &=&   \frac{m}{k_0} + \frac{im}{2L_0}\frac{1-e^{-ik_0L_0}}{k_0^2} - \frac{im}{2L_0}\frac{1-e^{ik_0L_0}}{k_0^2} \nonumber\\
&=& \frac{m}{k_0} - \frac{m}{k_0} \frac{\sin(k_0L_0)}{k_0 L_0}.
\end{eqnarray}

Next we evaluate the integral 
\begin{eqnarray} \label{ADtau2}
  \int_{-\infty}^\infty \frac{dk}{2\pi} |f(k-k_0)|^2     \tau_{\rm ph}(k) 
 \end{eqnarray}
in Eq.~(\ref{defttun}). We follow the same procedure as for the calculation of $v^{-1}$ outlined above. Taking the double pole at $k=k_0 + i \eps$ coming from $|f(k-k_0)|^2$ and the poles at $k=\pm i\eps$ coming from the derivative with respect to the energy in  
\begin{eqnarray} 
\tau_{\rm ph}(k) = \frac{\pard \th(k)}{\pard E_k} = \frac{m}{2}\left(\frac{1}{k+i\eps}+ \frac{1}{k-i\eps}\right)  \frac{\pard \th(k)}{\pard k},
 \end{eqnarray}
 we obtain  (with $\theta'(k)= \pard\th(k)/\pard k$)
\begin{eqnarray} \label{ADtau3}
&&  \int_{-\infty}^\infty \frac{dk}{2\pi} |f(k-k_0)|^2     \tau_{\rm ph}(k) \\
&=&  \frac{m}{k_0} \th'(k_0) -  \frac{m}{k_0} \frac{\sin(k_0L_0)}{k_0 L_0} \th'(0) + O(1/L_0) \nonumber\\
&=&  \tau_{\rm ph}(k_0) -   \frac{\sin(k_0L_0)}{k_0^2 L_0} [k \tau_{\rm ph}(k)]_{k = 0} + O(1/L_0), \nonumber
 \end{eqnarray}
 where we neglected the residues at the poles of the phase time, which are also poles of the scattering amplitudes. As discussed in Appendix \ref{AppSB}, as long as the width and height of the barrier are non-zero, these poles give $O(1/L_0)$ contributions, 
which we neglect for large $L_0$. (Note that the term involving $\sin(k_0 L_0)$ is non-negligible when $k_0\sim 1/L_0$.) Inserting Eq.~(\ref{ADtau3})  into Eq.~(\ref{ttunredef}) we obtain, 
\begin{eqnarray} \label{DADB}
 \Delta\tau_A &=&   
 \tau_{\rm ph}(k_0) 
- \frac{\sin(k_0L_0)}{k_0^2 L_0} \left[k \tau_{\rm ph}(k)\right]_{k = 0}   - a v^{-1}.\nonumber\\
 \end{eqnarray}

 For $\Delta\tau_B$ we follow a similar procedure (see Appendix \ref{AppDADB}). The result is  
\begin{eqnarray} 
\label{DADB'}
\Delta\tau_B= \frac{m}{k_0} L_0\left[1-\left(\frac{\sin(k_0L_0/2)}{k_0L_0/2}\right)^2\right] - L_0v^{-1}.
\end{eqnarray}
The term $mL_0/k_0$ gives the time it takes the particle to travel the distance $L_0/2$ on each side of the barrier  with a speed $m/k_0$. 

The age difference is 
\begin{eqnarray}  
 t_{\psi_2,\psi_1} &=& t_{\psi_2,\psi_1}^{(0)}+\Delta\tau_A+\Delta\tau_B \nonumber \\
 &=& \tau_{\rm ph}(k_0) - \frac{\sin(k_0L_0)}{k_0^2 L_0}
\left[k \tau_{\rm ph}(k)\right]_{k= 0} \nonumber\\
&+& \frac{m}{k_0} L_0\left[1-\left(\frac{\sin(k_0L_0/2)}{k_0L_0/2}\right)^2\right] \label{adfinal}.
\end{eqnarray}
This can be written as 
\begin{eqnarray}  
 t_{\psi_2,\psi_1} &=& t_{\rm tunnel} + t_{\rm outside}, \label{adfinal2}
\end{eqnarray}
where
\begin{eqnarray} \label{DADB2}
t_{\rm tunnel} &=& av^{-1} + \Delta\tau_A\nonumber\\
&=& \tau_{\rm ph}(k_0) - \frac{\sin(k_0L_0)}{k_0^2 L_0}
\left[k \tau_{\rm ph}(k)\right]_{k= 0}.
\end{eqnarray}
and 
\begin{eqnarray} \label{tout}
t_{\rm outside} &=& L_0 v^{-1} + \Delta\tau_B \nonumber\\
&=& \frac{m}{k_0} L_0\left[1-\left(\frac{\sin(k_0L_0/2)}{k_0L_0/2}\right)^2\right]
\end{eqnarray}
is the time that the particle spends outside the barrier.

\section{Branch-point contribution} 
\label{BPFX}

In the previous section we evaluated the integrals involved in the age difference, by taking residues at the poles, including the poles $k=\pm i\eps$ (with $\eps\to 0$). These poles are in fact associated with a branch point of the energy.  Indeed, the energy of the particle is $E_k =  k^2/(2m)$  outside the potential barrier. For the dispersion $E_k\propto k^2$, the complex energy plane has two Riemann sheets $E_k=k^2/2m$ with $\mathop{\mathrm{Im}}k>0$ and with $\mathop{\mathrm{Im}}k<0$. We have the branch cut on $E_k>0$ and the branch point at $E_k=0$, or at $k=0$. For this reason we will call the residues at $k=\pm i\eps$ the branch-point contribution.

The terms due to the branch point are the terms containing the sine function in Eqs.~(\ref{adfinal}), (\ref{DADB2}), and~(\ref{tout}). These terms  vanish when
$k_0 L_0 \gg 1$ but are non-negligible when $k_0 L_0 \sim 1$.  Since we are considering large wave packets, this means that the branch-point effect appears when the momentum of the particle is close to zero ($k_0\sim 1/L_0$). 

As $k_0\to 0$ and $L_0\to \infty$ the branch-point terms  approach either zero or infinity depending on the limiting order. In the 
limit $\lim_{k_0\to 0} \lim_{L_0\to\infty}$ the branch-point terms vanish, but in the limit $\lim_{L_0\to\infty}\lim_{k_0\to 0}$ they diverge. This  extreme dependence on $L_0$ means that the tunneling time has no characteristic scale near the branch point. It depends on the size of the incoming wave packet rather than any intrinsic time scale associated with the barrier. Because of this  the tunneling time is not intrinsic around 
the branch point. This gives a negative answer to the question of the existence of an intrinsic tunneling time.
As mentioned in Introduction, the branch-point contribution is 
associated with the non-Markovian dynamics (\textit{i.e.}, power-law decay) with no 
characteristic time or length scales~\cite{Khalfin,Misra,Petrosky}.  

We may understand the physical origin of the branch-point contribution as follows. The particle states that we are considering are, in position representation,  truncated plane waves of large width $L_0$. In contrast, in momentum representation, these wave packets are high, narrow peaks centered at $k_0$ with a width or order $1/L_0$. This width expresses the  uncertainty in the momentum of the particle, which occur because the initial and final states are not plane waves but truncated plane waves. When $k_0$ approaches zero with $k_0 \lesssim 1/L_0$, $k_0$ eventually becomes smaller than the uncertainty range. As a result, it becomes increasingly likely for the particle to have negative momenta. Negative momenta have the following effect on the age difference between two states: if two states $\psi_1$ and $\psi_2$ have negative momenta only, and $\psi_2$ is located to the right of $\psi_1$, then $\psi_2$ is actually ``younger'' than $\psi_1$. Therefore, negative momenta give negative contributions to the age difference between $\psi_2$ and $\psi_1$. On the contrary, if the momenta are positive, then the age difference is positive.   As a result,  as $k_0\to 0$, the average age difference, including both negative and positive age differences, tends to zero. This can be verified by taking this limit in Eq.~(\ref{adfinal}).

In short, the branch-point terms in the age difference, Eq.~(\ref{adfinal}),  express a reverse flow of the particle due to momentum fluctuations rooted in the uncertainty principle. The term involving $\sin(k_0 L_0)$ represents the reverse flow through the potential barrier, while the term involving $\sin^2(k_0 L_0/2)$ represents the reverse flow outside the barrier. 

The branch-point terms give negative contributions to the age difference, and hence they decrease the average time it takes the particle to move from its initial state $\psi_1$ to its final state $\psi_2$. In a sense, the branch-point causes the particle to speed up when it has a very small classical velocity.

\section{Resonance contribution}

In this section we will isolate the contribution to the tunneling time coming from the resonances of the scattering amplitudes over the barrier. We will show that, in contrast to the branch-point contribution,  this contribution is an intrinsic function of $E_0$, being independent of the size of the wave packet. 
 Moreover, we will show that the resonance contribution  can be written as a weighted sum of lifetimes associated with each resonance of the scattering amplitudes.  

The resonances appear in $ \Delta\tau_A$;
\begin{eqnarray}
 \Delta\tau_A &=&      -\frac{1}{2} \sum_{\alpha=\pm}  \int_{-\infty}^\infty \frac{dk}{2\pi}   |f(k-k_0)|^2  \frac{1}{F_\alpha} i \frac{\pard}{\pard E_k}  F_\alpha \nonumber\\
&=&     -\frac{1}{2} \sum_{\alpha=\pm}  \int_{-\infty}^\infty \frac{dk}{2\pi}   |f(k-k_0)|^2  i \frac{\pard}{\pard E_k} \ln\, F_\alpha.
\qquad
 \end{eqnarray}
Taking the pole at $k=k_0$, we obtain the residue
\begin{eqnarray}\label{AFlog}
 \Delta\tau_{k_0} =   -\frac{1}{2} \sum_{\alpha=\pm}  i \frac{\pard}{\pard E_{k_0}} \ln\,F_\alpha(k_0).
 \end{eqnarray}
 We write the scattering amplitudes as functions of the energy $E_0 = k_0^2/(2m)$ in the form
\begin{eqnarray} \label{AFEpole}
 F_\pm(E_0) = \prod_j \frac{E_0-(E_j^\pm)^*}{E_0-E_j^\pm}G_{\pm}(E_0),
 \end{eqnarray}
 where each $ F_\pm(E_0)$ has two branches, since $E_0 = \left(\pm k_0\right)^2/\left(2m\right)$. The product includes all the resonance poles $E_j^\pm$ of the scattering 
amplitude. The complex conjugate 
resonance $(E_j^\pm)^*$ must appear in the numerator because $F_\pm$ has modulo 
$1$. The 
function $G_{\pm}$ is a remainder factor with modulo $1$ as well. 

Using Eq.~(\ref{AFEpole})  we obtain from Eq.~(\ref{AFlog})
\begin{eqnarray}
\label{Lorentzian}
 \Delta\tau_{k_0} &=&   -\frac{1}{2} \sum_{\alpha=\pm}\left[ \sum_j  \left(\frac{i}{E_0-(E_j^\alpha)^*} -  \frac{i}{E_0-E_j^\alpha}\right)\right.\nonumber\\
 &+&\left.  i\frac{\pard}{\pard E_{k_0}} \ln G_{\alpha}(E_0) \right].
 \end{eqnarray}
By writing the real and imaginary parts of the poles as 
\begin{equation}
E_j^\pm = E_{Rj}^\pm - i\frac{\Gamma_j^\pm}{2},
\end{equation}
we obtain
\begin{eqnarray} \label{AFEpole2}
\Delta\tau_{k_0}&=& 2\sum_{\alpha=\pm}\sum_j \frac{1}{\Gamma_j^\alpha} \frac{(\Gamma_j^\alpha/2)^2}{(E_0-E_{jR}^\alpha)^2 + (\Gamma_j^\alpha/2)^2}\nonumber\\
&-&\frac{i}{2} \sum_\pm \frac{\pard}{\pard E_{k_0}} \ln G_{\alpha}(E_0).
 \end{eqnarray}
 The first term in the right-hand side is  the resonance-pole contribution to the tunneling time. This is the weighted average of the resonance lifetime $1/\Gamma_j^\pm$ with the Lorentzian weights between 0 and 1.
The factor 2  in front of the summations takes account of the fact that the particle comes into and goes out of the barrier to tunnel, while the resonance lifetime is only the time it takes the particle to go out of the scattering potential.

\section {Results for   a square barrier} 
\label{Results}
In this section we will discuss the behavior of the tunneling time and age difference as functions of the momentum of the particle $k_0$
for a square barrier.  We have 
\begin{eqnarray}
V(x) =
\left\{ \begin{array}{ll}
        V , & |x|\le a/2,\\
        0,  &   |x|>a/2,\\
        \end{array} \right.
 \end{eqnarray}

\begin{figure}[t]
\includegraphics[width=\columnwidth]{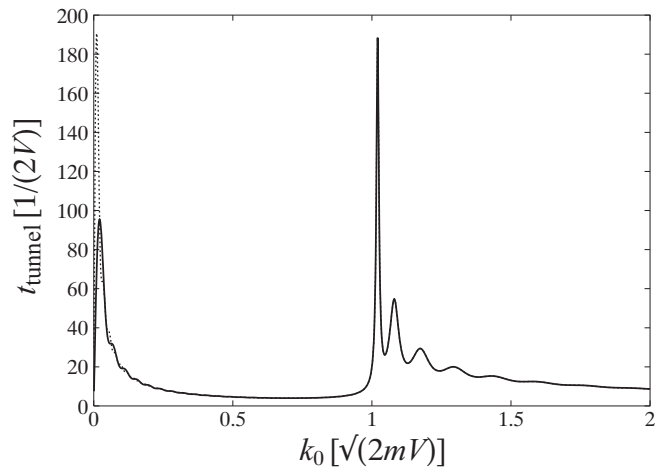}
\caption{The tunneling time $t_{\rm tunnel}$ \textit{vs}.\ the momentum of the particle $k_0$, for $a=15$, $m=1$, $2mV=1$, $L_0=150$ (solid),  and  $L_0=300$ (dotted).} 
\label{ftau2}
\end{figure}

The phase time is given \cite{Olkhovsky} by
\begin{eqnarray}\label{tphsq}
\tau_{\rm ph}(k) =  \frac{m}{k} \frac{1}{\kappa} \frac{l_0^4 \sinh(2\kappa a) + 2 a\kappa k^2 (\kappa^2-k^2)}
{l_0^4 \sinh^2(\kappa a) + 4 \kappa^2 k^2},
 \end{eqnarray}
where $l_0^2 = \kappa^2 + k^2 = 2mV$. Using Eqs.~(\ref{DADB2}) and~(\ref{tphsq}) we obtained the numerical plot of the tunneling time in 
Fig.~\ref{ftau2}  for two different widths of the incoming wave packet ($L_0=150$  and $L_0=300$), which are large compared to the width of the barrier ($a=15$).
As 
can be seen, the tunneling time has peaks near the  resonance poles of the 
scattering amplitudes $F_\pm$ (around $k_0=1.1$). According to our theoretical 
analysis, in this region the tunneling time should be independent of $L_0$ for 
large $L_0$, which is confirmed by the numerical plot.  Thus $t_{\rm tunnel}$ is 
an intrinsic function of $k_0$ in the resonance region.
 
On the other hand, near the  branch point,  \textit{i.e.}, near 
$k_0=0$, the tunneling time depends on the size $L_0$ of the incoming wave 
packet.  This means that there is no unique universal tunneling  time. 

\begin{figure}[t]
\includegraphics[width=\columnwidth]{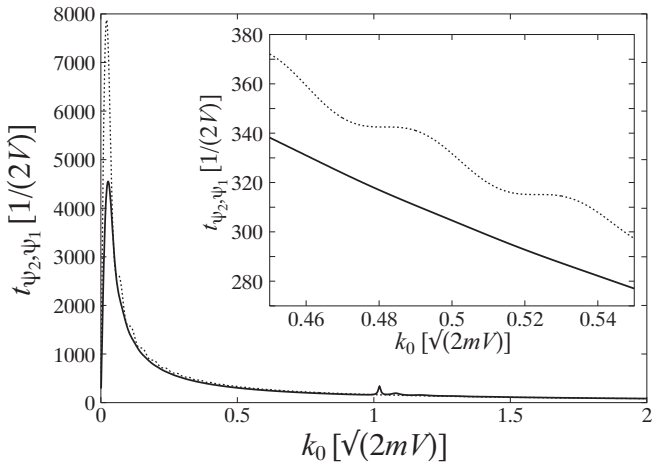}
\caption{{\it Main plot:}  The age difference $t_{\psi_2,\psi_1}$ (solid),  
and the age difference with no barrier $t_{\psi_2,\psi_1}^{(0)}$ (dashed) \textit{vs}.\ $k_0$, for $L_0=150$, $a=15$, $m=1$, and  
$2mV=1$. {\it Inset:} A zoom-in view around $k_0=0.5$.} 
\label{ftau3}
\end{figure}

Figure~\ref{ftau3}  compares the age difference $t_{\psi_2,\psi_1}$ with the barrier present  and the age difference $t_{\psi_2,\psi_1}^{(0)}$ without the barrier.
Note that around $k_0=0.5$ (inset), $t_{\psi_2,\psi_1} < t_{\psi_2,\psi_1}^{(0)}$. The particle is in a sense accelerated by the barrier. This 
 is  the Hartman effect~\cite{Hartman,Nimtz,Oetal,Olkhovsky,Jakiel}.  Close to $k_0=0$, $t_{\psi_2,\psi_1}$ is also smaller than  $t_{\psi_2,\psi_1}^{(0)}$. This  effect is not related to the Hartman effect. It is due to the branch-point effect, that is, due to the reverse flow of the particle caused by momentum fluctuations, as we discussed at the end of Section \ref{BPFX}. When $k_0$ is close to zero this reverse flow is enhanced by the reflection due to the barrier. This, on average,  makes the particle arrive sooner when the barrier is present than when there is no barrier.

\section{Concluding remarks}

We have proposed a definition of tunneling time obtained from the change of expectation value of the time operator. We considered spatially large incoming and outgoing wave packets (truncated plane waves with average momentum $k_0$).  Our tunneling time is the average phase time, averaged over the momemtum distribution of the incident particle. It reduces to the phase time $\tau_{\rm ph}(k_0)$ when the energy of the tunneling particle is far from the branch point of the energy continuum. 

However, near the branch point we obtain a deviation from the phase time $\tau_{\rm ph}(k_0)$; the deviation depends on the size of the incoming and outgoing wave packets. This deviation gives a non-intrinsic character to the tunneling time.  
It may be interesting to see if a tunneling experiment of slow 
particles (with $k_0\approx 1/L_0$) shows dependence of the tunneling time 
on the size of the wave packet  as predicted here.   

Our calculations have centered on a symmetrical barrier, and symmetrical initial and final states. One could consider asymmetric configurations. Another possible extension of our work is to consider the case where the wave packets have positive-momentum components only. In this case we expect that there will still be branch-point effects, but they will take a form different from the one discussed in this paper. 

The definition of tunneling time we presented here is by no means the only possible definition. There are many other definitions, and one might wonder if our main prediction, the appearance of deviations from the phase time for slow particles, is not an artifact of our definition. 
Again, it will be important to detect the deviations in a tunneling experiment.

Olkhovsky and Recami \cite{OR07} have argued that the domain of the time operator should be restricted to functions of the energy $E$ that vanish at $E=0$ (the branch point). In this way the time operator becomes Hermitian and has real expectation values. Moreover, even if the domain includes  functions that do not vanish at $E=0$, they have proposed to use  a bilinear time operator, which, again gives real expectation values. In our approach, however, this issue  is not very relevant because the age difference that we obtained is real to begin with. Using the bilinear time operator proposed by Olkhovsky and Recami  \cite{OR07} instead of the operator in Eq.~(\ref{timeop}) will give the same  age difference~(\ref{agediff}). 
 
As mentioned above, our tunneling time is the average phase time. The phase time has found some experimental support as a good measure of the tunneling time \cite{Jakiel}, (although it is not universally agreed that the phase time is the correct tunneling time). New experiments  (and possibly more theoretical work) could confirm or negate our prediction. 

\acknowledgments
We thank Prof.~T.~Petrosky, Dr.~Sungyun Kim, Dr.~M.~Machida, Dr.~K.~Sasada and Mr.~Y.~Aoki
for helpful suggestions.
G.O.\ thanks  the Institute of Industrial Science and the Hatano Lab.\ for their 
hospitality and support. N.H.\ acknowledges support by Grant-in-Aid for 
Scientific 
Research (No.~17340115) from the Ministry of Education, Culture, Sports, Science 
and Technology and by CREST of Japan Science and Technology Agency.
\bigskip

\appendix
\section{Calculation of the age difference}
\label{AppB}
Here we will show the main steps involved in the calculation of the age difference between the final and initial states $\psi_2$ and $\psi_1$. Defining
\begin{eqnarray} \label{Afg}
\bra k|\psi_1\ket &=& f^*(k-k_0), \\
\bra k|\psi_2\ket &=&  g^*(k-k_0) f^*(k-k_0),
\end{eqnarray}
where 
\begin{eqnarray} \label{Afksq}
f^*(k-k_0) = \frac{1}{\sqrt{L_0}}e^{i(k-k_0)a/2} \frac{1-e^{i(k-k_0)L_0}}{-i(k-k_0)}
 \end{eqnarray}
 and
\begin{eqnarray}\label{Agcck}
g^*(k-k_0) = e^{-i(k-k_0)(L_0+a)},
\end{eqnarray}
we have
\begin{eqnarray}
\bra\psi_1|E_k^\pm \ket &=& \frac{1}{2} \left(  f(k-k_0) + F_\pm(k)  f(-k-k_0) \right) \nonumber\\
\bra\psi_2|E_k^\pm \ket &=& \pm\frac{1}{2} \left(  f(-k-k_0)g(-k-k_0) \right. \nonumber\\
&+& \left.  F_\pm(k) f(k-k_0)g(k-k_0) \right)
\end{eqnarray}
and (see Eq.~({\ref{psitpsi}))
\begin{eqnarray}\label{A1t1}
\bra\psi_1|\tha|\psi_1\ket &=&   \frac{i}{4} \sum_{\alpha=\pm}  \int  \frac{dk}{2\pi}  \\
&\times&\left(  f(k-k_0) \frac{\pard}{\pard E_k} f^*(k-k_0)  \right.
\nonumber\\
&&+ F_\alpha(k) f(-k-k_0)  \frac{\pard}{\pard E_k} F_\alpha^*(k)f^*(-k-k_0) \nonumber\\
&&+  F_\alpha (k) f(-k-k_0) \frac{\pard}{\pard E_k} f^*(k-k_0) \nonumber\\
&&+ \left.f(k-k_0) \frac{\pard}{ E_k} F_\alpha^*(k)f^*(-k-k_0)\right) \nonumber
\end{eqnarray}
as well as
\widetext
\begin{eqnarray}\label{A2t2}
&& \bra\psi_2|\tha|\psi_2\ket =   \frac{i}{4} \sum_{\alpha=\pm}  \int  \frac{dk}{2\pi}  \\
&\times&\left(  F_\alpha(k) f(k-k_0) g(k-k_0)  \frac{\pard}{\pard E_k} F_\alpha^*(k) g^*(k-k_0) f^*(k-k_0)  \right.\nonumber\\
&+&  \left. f(-k-k_0)g(-k-k_0)  \frac{\pard}{\pard E_k} f^*(-k-k_0) g^*(-k-k_0) \right. \nonumber\\
&+& \left. F_\alpha(k) f(k-k_0) g(k-k_0) \frac{\pard}{\pard E_k} f^*(-k-k_0) g^*(-k-k_0) \right.\nonumber\\
&+& \left.  f(-k-k_0)g(-k-k_0) \frac{\pard}{\pard E_k} F_\alpha^*(k)f^*(k-k_0) g^*(k-k_0)\right). \nonumber 
\end{eqnarray}
\endwidetext
We first consider the first and second terms  in the right-hand side of Eqs.~(\ref{A1t1}) and~(\ref{A2t2}). Using the relations $|F_\pm(k)|^2 = 1$ and $|g(k)|^2=1$ for real $k$, we find that they give the following contribution to the age difference:
\begin{eqnarray}
&& t_{\psi_2,\psi_1}^{\rm (A)} =    \frac{i}{4} \sum_{\alpha=\pm}  \int \frac{dk}{2\pi} \\
&\times& \left( |f(k-k_0)|^2 g(k-k_0) \frac{\pard}{\pard E_k} g^*(k-k_0) \right. \nonumber\\
&&+ |f(-k-k_0)|^2 g(-k-k_0) \frac{\pard}{\pard E_k} g^*(-k-k_0) \nonumber\\
&&+   |f(k-k_0)|^2  \frac{1}{F_\alpha^*} \frac{\pard}{\pard E_k}  F_\alpha^* \nonumber\\
&&-  \left.|f(-k-k_0)|^2  \frac{1}{F_\alpha^*} \frac{\pard}{\pard E_k}  F_\alpha^* \right).\nonumber
\end{eqnarray}
For the terms involving $-k$ we change the variable of integration from $k$ to $-k$. Noting that 
$F^*_\pm(-k) = F_\pm(k)$ we obtain 
\begin{eqnarray}\label{Aaged}
\lefteqn{
t_{\psi_2,\psi_1}^{\rm(A)} =    \int \frac{dk}{2\pi}  |f(k-k_0)|^2 g(k-k_0) i \frac{\pard}{\pard E_k} g^*(k-k_0)
} \nonumber\\ 
&-&   \frac{1}{4} \sum_{\alpha=\pm}  \int \frac{dk}{2\pi} \left( |f(k-k_0)|^2  \frac{1}{F_\alpha} i \frac{\pard}{\pard E_k}  F_\alpha +  \rm{c.c.}  \right),
\nonumber\\
\end{eqnarray}
where c.c.\ denotes complex conjugate. The first term in the right-hand side of Eq.~(\ref{Aaged}) does not involve the potential barrier. It therefore gives the age difference with no barrier, 
\begin{eqnarray}\label{Atprop}
\lefteqn{ 
t_{\psi_2,\psi_1}^{(0)}
} \nonumber\\
&=&      \int \frac{dk}{2\pi}  |f(k-k_0)|^2 g(k-k_0) i \frac{\pard}{\pard E_k} g^*(k-k_0) \nonumber\\
&=&      (L_0+a) \int \frac{dk}{2\pi}  |f(k-k_0)|^2 \frac{\pard k}{\pard E_k} \nonumber\\
&=&      (L_0+a) v^{-1},
\end{eqnarray}
 where 
 \begin{eqnarray}\label{Avav}
v^{-1}  =      \int_{-\infty}^\infty \frac{dk}{2\pi}  |f(k-k_0)|^2  \frac{\pard k}{\pard E_k}
\end{eqnarray}
is the average inverse group velocity. On the other hand, the second term in the r.h.s.\ of Eq.~(\ref{Aaged}) may be written as
\begin{eqnarray}\label{Attun'}
 \Delta\tau_A =      -\frac{1}{2} \sum_{\alpha=\pm}  \int \frac{dk}{2\pi}   |f(k-k_0)|^2  \frac{1}{F_\alpha} i \frac{\pard}{\pard E_k}  F_\alpha
 \end{eqnarray}
 because $F_\pm$ has modulo $1$ and we have $F^*_\pm = 1/F_\pm$. When there is no potential barrier, we have $F_\pm(k) = \pm 1$. Hence $\Delta\tau_A$ vanishes when there is no barrier; it is a correction to the age difference due to the barrier. 

For the  third and fourth terms in Eqs.~(\ref{A1t1}) and~(\ref{A2t2}), we use the relation
\begin{eqnarray}
f(\kappa) g (\kappa) = f^*(\kappa).
\end{eqnarray}
 The contribution to the age difference from  the  third and fourth terms in Eqs.~(\ref{A1t1}) and~(\ref{A2t2}) is then
\begin{eqnarray}
&& t_{\psi_2,\psi_1}^{\rm (B)} =    \frac{i}{4} \sum_{\alpha=\pm}  \int \frac{dk}{2\pi} \\
&\times& \left(F_\alpha(k) f^*(k-k_0) \frac{\pard}{\pard E_k} f(-k-k_0)\right. \nonumber\\
&&+ \left.  f^*(-k-k_0) \frac{\pard}{\pard E_k} F^*_\alpha(k) f(k-k_0)\right. \nonumber\\
&&-  \left. F_\alpha(k) f(-k-k_0) \frac{\pard}{\pard E_k} f^*(k-k_0)\right. \nonumber\\
&&-   \left. f(k-k_0) \frac{\pard}{\pard E_k} F^*_\alpha(k) f^*(-k-k_0)\right).
\end{eqnarray}
Changing $k\to-k$ in the second and fourth terms in the above expression and using $F_\pm^*(k) = F_\pm(-k)$ as well as $f^*(k-k_0) = f(-k+k_0)$, we obtain 
\begin{eqnarray} \label{Anewbr2}
&& t_{\psi_2,\psi_1}^{\rm (B)} = \Delta\tau_B   = \frac{i}{2} \sum_{\alpha=\pm}  \int \frac{dk}{2\pi} \nonumber\\
&\times& \left(F_\alpha(k) f(-k+k_0) \frac{\pard}{\pard E_k} f(-k-k_0)\right. \nonumber\\
&&-  \left. F_\alpha(k) f(-k-k_0) \frac{\pard}{\pard E_k} f(-k+k_0)\right).
\end{eqnarray}
When there is no potential barrier,  $ t_{\psi_2,\psi_1}^{\rm (B)}$ vanishes because $F_\pm(k) = \pm 1$ with no barrier. Therefore, $ t_{\psi_2,\psi_1}^{\rm (B)}$ is another correction to the age difference coming from the barrier and we write it as $ \Delta\tau_B$. The  total age difference is 
\begin{equation}
t_{\psi_2,\psi_1} = t_{\psi_2,\psi_1}^{(0)} +  \Delta\tau_A + \Delta\tau_B.
\end{equation}
 
 \section{Scattering amplitude for a square barrier}
\label{AppSB}
In this appendix we will write down the scattering amplitudes for a square barrier potential, and we will consider its limits when the momentum of the particle goes to zero and the width or height of the barrier go to zero, with the aim of establishing a range of validity of the age difference calculated in the text.

The scattering amplitudes are  obtained from their definitions $F_\pm(k) = \exp(-ika)(R(k)\pm T(k))$ and the well-known expressions for the reflection coefficient $R(k)$ and the transmission coefficient $T(k)$. The result is  
\begin{equation}\label{AFpmsq}
F_\pm(k) = e^{-ika} \frac{(1\pm e^{-\kappa a}) k -i(1 \mp e^{-\kappa a})\kappa}{(1\pm e^{-\kappa a}) k + i(1 \mp e^{-\kappa a})\kappa},
 \end{equation}
where   $V$ is the height of the barrier, $a$ is the width of the barrier, and $\kappa =  \sqrt{2mV-k^2}$. 

We will consider first the limit $a\to 0$. 
It turns out that the limits $k\to 0$ and $a\to 0$ of the scattering amplitude $F_+$ are not interchangeable. Indeed, when $k\to 0$  we have  $\kappa a = a \sqrt{2mV}$. Thus, for any $V>0$ and $a>0$, we have  $\lim_{k\to 0} F_\pm(k) = -1$.  However, when $a\to 0$ with $k>0$ we have $F_\pm(k) \to \pm 1$. We will discuss next how this affects the age difference.

The integrands that appear in the age difference are peaked around $k=k_0$; therefore in the following discussion we will consider $k\sim k_0$. 
We will focus on the branch-point effect, which  appears when $k \sim k_0\sim 1/L_0 \sim 0$. Moreover, we will consider the expression for the amplitudes when $a$ is small, so that $\kappa a \ll 1$, or 
\begin{equation} \label{mVa}
\sqrt{mV} a \ll 1.
 \end{equation}
Then we have
\begin{equation}\label{AFpmsq2}
F_+(k) = \frac{k + i (mV - k^2/2)a}{k - i (mV - k^2/2)a},
 \end{equation}
\begin{equation}\label{AFpmsq3}
F_-(k) = \frac{k -2i \sqrt{2mV - k^2}/a}{k + 2i \sqrt{2mV - k^2}/a}.
 \end{equation}
These expressions show that for $F_-$ the limits $k\to 0$ and $a\to 0$ are interchangeable. For either limit we have $F_- \to -1$. However, for $F_+$ the limits are not interchangeable, as mentioned earlier.  When the limits are taken so that $amV \gg k \sim 1/L_0$, then we have 
\begin{equation}\label{AFpmsq4}
\lim_{a\to 0} \lim_{k\to 0} F_+ (k) = -1,
 \end{equation}
but when $amV \ll k \sim 1/L_0$, then we have
\begin{equation}\label{AFpmsq5}
\lim_{k\to 0}  \lim_{a\to 0} F_+ (k) =  1.
 \end{equation}
This   gives  a discontinuity between the age difference with the barrier present (Eq.~(\ref{adfinal})) and the age difference with no barrier (Eq.~(\ref{tprop})) when we take the limit $a\to 0$ in the former. This discontinuity appears because when we derived Eq.~(\ref{adfinal}) we neglected the pole contributions coming from the scattering amplitudes, arguing that they gave $O(1/L_0)$ corrections. This was fine as long as the width $a$ of the barrier was finite. However, when $a\to 0$, the poles of the scattering amplitudes give terms comparable to the term coming from the poles at $k=\pm i\eps$ (the branch-point contributions). 

Specifically, one can see that when $a\to 0$ the scattering amplitude $F_+(k)$ in Eq.~(\ref{AFpmsq2}) has a pole at $k = imVa$. When $a\to 0$ this is essentially a pole at the branch point, giving a non-negligible residue. If we include this residue, the discontinuity mentioned above is removed. 
This brings the question of how large $a$ has to be so that Eq.~(\ref{adfinal}) is valid. 

When $a\to 0$, the residue of the pole at $k = imVa$ involves the term 
\begin{equation}\label{AFpmsq6}
\exp(ikL_0) = \exp(-mVaL_0)
 \end{equation}
 coming from the incoming or outgoing wave-functions (see Eq.~(\ref{fksq})).  This term vanishes if
\begin{equation}\label{mVa2}
a\gg 1/(mV L_0).
 \end{equation}
Therefore, Eq.~(\ref{adfinal}) is valid only if Eq.~(\ref{mVa2}) holds. This condition is consistent with the condition we mentioned earlier above Eq.~(\ref{AFpmsq4}).

Similar arguments apply when we take the limit $V\to 0$ instead of $a\to 0$. Equation (\ref{adfinal}) is valid  if Eq.~(\ref{mVa2}) holds, \textit{i.e.}, if  $V\gg 1/(ma L_0)$.

 \section{Explicit form of  $\Delta \tau_B$}
\label{AppDADB}
 
 In this appendix we will evaluate the term $\Delta \tau_B$ in Eq.~(\ref{ttun'b}). Writing the derivative $\pard /\pard E_k$ in terms of $k$, we have (see Eq.~(\ref{Anewbr2}))
\begin{eqnarray} \label{ADTB'}
&&\Delta\tau_B   = \frac{i}{2} \sum_{\alpha=\pm}  \int \frac{dk}{2\pi} \frac{m}{k} \nonumber\\
&\times& \left(F_\alpha(k) f(-k+k_0) \frac{\pard}{\pard k} f(-k-k_0)\right. \nonumber\\
&&-  \left. F_\alpha(k) f(-k-k_0) \frac{\pard}{\pard k} f(-k+k_0)\right).
\end{eqnarray}
 To evaluate the integral we will  close the integration contour using either the upper or the lower infinite semicircles. The integrals are then reduced to summations over the residues of the  poles inside the contour. Since
\begin{eqnarray} \label{Afksq2}
f(-k\pm k_0) &=& \frac{1}{\sqrt{L_0}}e^{i(k\mp  k_0)a/2} \frac{1-e^{i(k\mp  k_0)L_0}}{-i(k\mp  k_0)},\nonumber\\
 \end{eqnarray}
 we will close the contour in the upper infinite semicircle.  The functions $f(-k\pm k_0)$ and their derivatives have no poles at $k=\pm k_0$.  The scattering amplitude $F_\pm(k)$ may have poles in the upper half-plane. However, any residues of these poles are, except for phase factors, independent of $L_0$. Therefore, due to the $1/\sqrt{L_0}$ factor in Eq.~(\ref{Afksq2}), the poles of the scattering amplitude give $O(1/L_0)$ contributions, which we neglect. 
 Finally, the $1/k$ factor   gives a branch-point contribution.  Similarly to Eq.~(\ref{Attunsq3'}) we interpret this factor as a principal part. 
Equation~(\ref{ADTB'}) takes the explicit form
\begin{eqnarray} \label{ADTB''}
\Delta\tau_B   = S(k_0)- S(-k_0),
\end{eqnarray}
 where
 \widetext
\begin{eqnarray} \label{ADTB3}
S(k_0) &=&  \frac{i}{2L_0} \sum_{\alpha=\pm}  \int \frac{dk}{2\pi} F_\alpha(k) \frac{m}{2}\left(\frac{1}{k-i\eps} + \frac{1}{k+i\eps}\right) \\
&\times& \left[e^{i(k-  k_0)a/2} \frac{1-e^{i(k-  k_0)L_0}}{-i(k-  k_0)}
 \right] \frac{\pard}{\pard k}  \left[e^{i(k+  k_0)a/2} \frac{1-e^{i(k+  k_0)L_0}}{-i(k+  k_0)}
 \right], \nonumber
\end{eqnarray}
 or
\begin{eqnarray} \label{ADTB4}
S(k_0) &=&  \frac{i}{2L_0} \sum_{\alpha=\pm}  \int \frac{dk}{2\pi} F_\alpha(k) \frac{m}{2}\left(\frac{1}{k-i\eps} + \frac{1}{k+i\eps}\right) \left[e^{i(k-  k_0)a/2} \frac{1-e^{i(k-  k_0)L_0}}{-i(k-  k_0)}
 \right] \\
&\times&  e^{i(k+  k_0)a/2} \left\{\left[ \frac{1-e^{i(k+  k_0)L_0}}{-i(k+  k_0)} \right]\left(\frac{ia}{2}-\frac{1}{k+k_0}\right) + \frac{-iL_0}{-i(k+k_0)} e^{i(k+  k_0)L_0} \right\}. \nonumber
\end{eqnarray}
\endwidetext \noindent Taking the residue at $k=i\eps$, we obtain 
\begin{eqnarray} 
\Delta\tau_B   = \frac{m}{2} \sum_{\alpha=\pm} F_\alpha(0)   \left[2\frac{1-\cos(k_0 L_0)}{k_0^3 L_0} - \frac{\sin(k_0 L_0)}{k_0^2}\right].\nonumber\\
\end{eqnarray}

When $k=0$, the transmission coefficient is $T=0$ and the reflection coefficient is $R=-1$. Therefore, the amplitudes $F_\pm(k) = (R(k)\pm T(k)) e^{ika}$ become $F_\pm(0) = -1 $ at $k=0$.  Hence we have
\begin{eqnarray}  \label{ADTB}
\Delta\tau_B   =  -  m \left[2 \frac{1-\cos(k_0 L_0)}{k_0^3 L_0}  -\frac{\sin(k_0 L_0)}{k_0^2}\right],
\end{eqnarray}
 or, using $\cos(x) = 1 - 2\sin^2(x/2)$ as well as Eq.~(\ref{Avinv}), we arrive at
 \begin{eqnarray}  \label{ADTB2}
 \Delta\tau_B= \frac{m}{k_0} L_0\left[1-\left(\frac{\sin(k_0L_0/2)}{k_0L_0/2}\right)^2\right] - L_0v^{-1}. 
  \end{eqnarray}



\end{document}